\documentclass[aps,prl,twocolumn,groupedaddress]{revtex4-1}
\usepackage{amsmath} 
\usepackage{graphicx}
\usepackage{dcolumn}
\usepackage{mathrsfs}
\usepackage{bm}
\usepackage{subfigure}
\usepackage{natbib}
\usepackage{chemfig}
\usepackage[colorlinks,linkcolor=blue,anchorcolor=blue,citecolor=blue,urlcolor=blue]{hyperref}

\begin{document}
\title{Microwave Quantum Illumination via Cavity Magnonics}
\author{Qizhi Cai$^{1,2}$}
\author{Jinkun Liao$^{1,2}$}\thanks{E-mail: jkliao@uestc.edu.cn}
\author{Bohai Shen$^{1}$}
\author{Guangcan Guo$^{1,3}$}
\author{Qiang Zhou$^{1,2,3,}$}\thanks{E-mail: zhouqiang@uestc.edu.cn}
\address{$^{1}$Institute of Fundamental and Frontier Sciences, University of Electronic Science and Technology of China, Chengdu, Sichuan, China}
\address{$^{2}$School of optoelectronic science and engineering, University of Electronic Science and Technology of China, Chengdu, Sichuan, China}
\address{$^{3}$CAS Key Laboratory of Quantum Information, University of Science and Technology of China, Hefei, China}

\date{\today}
\begin{abstract}
Quantum illumination (QI) is a quantum sensing protocol mainly for target detection which uses entangled signal-idler photon pairs to enhance the detection efficiency of low-reflectivity objects immersed in thermal noisy environments. Especially, due to the naturally occurring background radiation, the photon emitted toward potential targets more appropriately lies in the microwave region. Here, we propose a hybrid quantum source based on cavity magnonics for microwave QI, where the medium that bridges the optical and the microwave modes is magnon, the quanta of spin wave. Within experimentally accessible parameters, significant microwave-optical quantum resources of interest can be generated, leading to orders of magnitude lower detecting error probability compared with the electro-optomechanical prototype quantum radar and any classical microwave radar with equal transmitted energy.
\end{abstract}

\pacs{Valid PACS appear here}
\maketitle

\textit{Introduction.—}Quantum sensing is a rapidly developing research field emerged with other advanced quantum technologies \citep{ref1, ref2, ref3, ref4, ref5, ref6, ref7, ref8, ref9, ref10}, aiming to utilize the extreme sensitivity of quantum systems when influenced by external perturbations to develop sensors. With the help of quantum resources like entanglement \citep{ref11}, the performance of signal detection \citep{ref12}, phase estimation \citep{ref13}, and microscopy \citep{ref14} could be enhanced. However, among these applications, the loss of entanglement caused by environmental noise or decoherence often decreases or even destroys the quantum-enhanced sensitivity. Quantum illumination (QI) primarily for target detection is an exception \citep{ref15, ref16, ref17}. In the QI protocol, quantum correlated signal-idler photon pairs are used to improve the detection efficiency of low-reflectivity objects hidden in thermal noisy environments. Despite the entanglement-breaking channel, the surviving quantum correlation still enough to achieve the performance that any classical strategy cannot reach, promisingly extending the applications of quantum sensing such as noninvasive diagnostic scanner \citep{ref18} and quantum radar \citep{ref19, ref20}. In part due to the naturally-occurring background radiation and prevailing radar working wavelength, Barzanjeh $et$ $al$ proposed microwave QI \citep{ref21}, sending microwave modes toward target, with the help of microwave-optical entanglement source like electro-optomechanical (EOM) system. From Ref.\citep{ref21, ref22}, one may conclude that it is the quantum correlations normalized by per emitted microwave photon between the unlaunched signal and idler modes that determine the quantum-enhanced sensitivity, since QI’s advantage is computed within equal transmitted energy. In this regard, in order to improve the performance of QI, i.e., lower the detection error probability, enlarging the signal-idler quantum correlations normalized by per emitted microwave photon is one of the key requirements.
\begin{figure*}
\centering
	\includegraphics[width=0.8\linewidth]{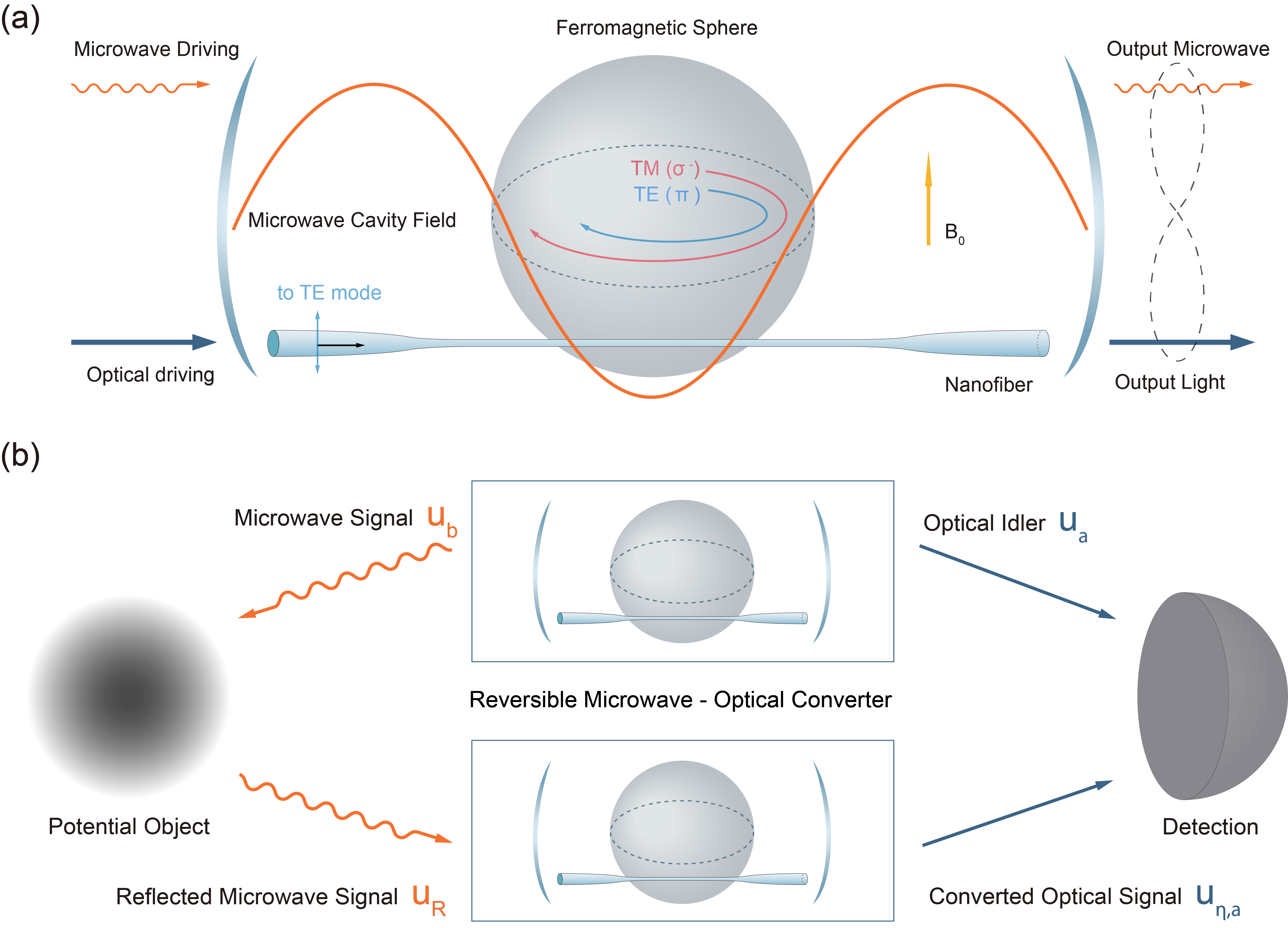}
	\caption{(a) Simple sketch of the magnonical reversible microwave-optical converter. A YIG sphere is placed in a microwave cavity and holds the magnon mode excited by an external bias magnetic field, which establishes the electromagnonical coupling via magnetic dipole interaction. The optical mode in nanofiber can evanescently couple to TE WGMs in the YIG sphere, which produces and establish the optomagnonical coupling between TM WGM and magnon mode. (b) Microwave-optical QI magnonical quantum source that generates quantum correlations between microwave-signal and optical-idler fields. The receiver collects the reflected microwave signal, upconverts it into optical domain meanwhile performs a phase-conjugate operation. Then, the converted signal together with the retained idler are fed into detector, whose measurement result decides the presence or absence of the target.}
	\label{fig:Fig1.png}
\end{figure*}

Here, we propose a scheme based on cavity magnonics \citep{ref23} to perform microwave QI that can fulfill the requirement mentioned above. In our proposal, the mediate that bridges the microwave and optical modes is magnon, the quanta of spin wave, which could couple and entangle with microwave photons in a strong-coupling regime \citep{ref24, ref25, ref26, ref27} as well as interact with optical photons via magnon-induced Brillouin scattering process \citep{ref28}, and also could serve to convert and entangle microwave and optical fields \citep{ref29, ref30}. Since the magnon's resonant frequency is $\sim$GHz, about three orders of magnitude higher than the mechanical resonator ($\sim$MHz) described in Ref.\citep{ref21}, the thermal excitation number governed by Planck's law is correspondingly smaller under the same temperature, leading to a smaller average photon number per emitted mode. Therefore, compared with the EOM strategy, 2 to 22 or more times of quantum resources per emitted microwave photon between output microwave-optical fields could be obtained, contributing to orders-of-magnitude lower detecting error probability in comparison with any classical microwave radar and EOM-prototype microwave quantum radar with equal transmitted energy. Furthermore, our proposal could also be realized via state-of-the-art technology, providing reachable potential for more applications in the quantum regime based on hybrid magnonical systems.

\textit{Description of the system.—}As depicted in Fig.1(a), we consider the magnon modes in yttrium iron garnet (YIG) sphere, excited by external bias magnetic field $B_0$, electromagnonically couples a microwave cavity mode via magnetic dipole interaction; meanwhile, the magnon modes share a three-wave process, the magnon-induced Brillouin scattering process, with optical TE and TM whispering gallery modes (WGMs) in the YIG sphere. The total Hamiltonian describing the system is $H = H_{0} + H_{int}$ where the free energy Hamiltonian is \citep{ref30}
\begin{equation}
\begin{split}
H_{0}=\sum\nolimits_{j}\hbar \omega _{aj} \hat{a}^{\dagger}_{j}\hat{a}_j+\hbar \omega _{m} \hat{m}^{\dagger}\hat{m}+\hbar \omega _{b} \hat{b}^{\dagger}\hat{b}, 
\end{split}
\end{equation}
and the interaction Hamiltonian read \citep{ref27, ref28, ref30}
\begin{equation}
\begin{split}
H_{int} = \hbar g_{ma}(\hat{a}_1 \hat{a}^{\dagger}_2 \hat{m}^{\dagger} + \hat{a}^{\dagger}_1 \hat{a}_2 \hat{m}) + \hbar g_{mb}(\hat{b}+\hat{b}^{\dagger})(\hat{m}+\hat{m}^{\dagger}),
\end{split}
\end{equation}
in which $\hat{a}_1$, $\hat{a}_2$, $\hat{b}$ and $\hat{m}$ are the annihilation operators of the $\pi$-polarized TE WGM, $\sigma^{-}$-polarized TM WGM, microwave and magnon modes, respectively, satisfying $[\hat{O}, \hat{O}^{\dagger}]=1$ ($O=a_j,b,m$). $\omega_{aj}$, $\omega_{b}$, and $\omega_m$ are optical, microwave, and magnon resonance frequency, while $g_{ma}$ and $g_{mb}$ are the optomagnonical and electromagnonical coupling rates, respectively.

We can linearize the optomagnonical interaction $H_{ma} = \hbar g_{ma}(\hat{a}_1 \hat{a}^{\dagger}_2 \hat{m}^{\dagger} + \hat{a}^{\dagger}_1 \hat{a}_2 \hat{m})$ by driving the TE mode $\hat{a}_1$ with a bright coherent tone, leading to $\hat{a}_1$ $\rightarrow$ $\alpha$. Then, the optomagnonical interaction simplifies to $H^{'}_{ma} = \hbar g_{ma} \alpha (\hat{a}^{\dagger}_2 \hat{m}^{\dagger} + \hat{a}_2 \hat{m})$, where we have chosen proper phase reference so that $\alpha$ can be taken real and positive. By applying the rotating-wave approximation, the electromagnonical interaction $\hbar g_{mb}(\hat{b}+\hat{b}^{\dagger})(\hat{m}+\hat{m}^{\dagger})$ could be linearized to $\hbar g_{mb}(\hat{b}\hat{m}^{\dagger}+\hat{b}^{\dagger}\hat{m})$ (valid when $\omega_b$, $\omega_m$ $\gg$ $g_{mb}$, $\kappa_b$, $\kappa_m$, where $\kappa_b$ and $\kappa_m$ are the damping rates of microwave and magnon mode, which is easily satisfied \citep{ref27, ref32, ref33}). Thereafter, with respect to the free Hamiltonian, we may write the linearized total Hamiltonian
\begin{equation}
\begin{split}
H^{'} = \hbar G_{ma} (\hat{a}^{\dagger} \hat{m}^{\dagger} + \hat{a} \hat{m}) +\hbar g_{mb}(\hat{b} \hat{m}^{\dagger}+\hat{b} ^{\dagger}\hat{m}),
\end{split}
\end{equation}
where we have denoted $\hat{a}_2$ as $\hat{a}$ and $G_{ma} \equiv g_{ma} \alpha$ for simplicity.

The first term in Eq.(3) shows that the optomagnonical interaction is a parametric-type interaction that could generate the entanglement between optical and magnon modes within the optomagnonical subsystem. The second term, the electromagnonical interaction, is the beam-splitter interaction that can transfer the optomagnonical entanglement into the microwave-optical intra-cavity entanglement. This intra-cavity entanglement could be transmitted to the entanglement between the propagating optical field $u_a$ and propagating microwave field $u_b$, as long as the optomagnonical rate $G_{ma}^2 / \kappa_a$ and electromagnonical rate $g^2 / \kappa_b$ exceed the decoherence rate of the magnon $\zeta = \kappa_m \bar{n}_b^T$, in which $\kappa_a$ is the damping rate of the TM WGM and $\bar{n}_b^T$ is the thermal excitation number of the magnon. These two conditions are important in QI's advantage as later will discuss.
\begin{figure*}
	\centering
	\includegraphics[width=1\linewidth]{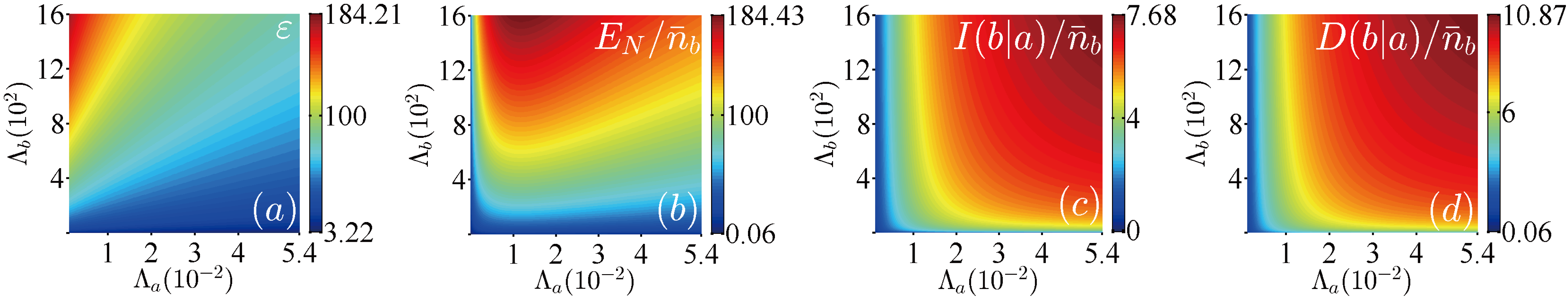}
	\caption{Quantum resources of magnonical QI source versus optical cooperativity $\Lambda_a$ and microwave cooperativity $\Lambda_b$: (a) metric of microwave-optical entanglement of propagating output fields, (b) the normalized logarithmic negativity $ E_N / \bar{n}_b $, (c) the normalized coherent information $ I(a|b) / \bar{n}_b $, and (d) the normalized quantum discord $ D(b|a) / \bar{n}_b $. The experimentally reachable parameters for magnon part: radius of the YIG sphere $r =$ 100 $\mu$m, external bias magnetic field $B_0 = 100$ mT, damping rate $\kappa_m / 2 \pi = $ 1 MHz; for optical part: optical quality factor $Q = 3 \times 10^{6}$, pump power $P_p =$ 60 mW, pump wavelength $\lambda_p = $ 1550 nm; for microwave part: resonant frequency $\omega_b = $ 9 GHz, damping rate $\kappa_b / 2 \pi = $ 1 MHz. The environmental temperature $T_{\textup{env}} = 30$ mK.
	}
	\label{fig:Fig2.png}
\end{figure*}

\textit{Microwave-optical quantum resources.—}The output propagating modes can be represented by the intra-cavity quantum noise operators, $\hat{a}_{in}$ for optical mode, $\hat{b}_{in}$ for microwave mode and $\hat{m}_{in}$ for magnon mode, via \citep{ref34}
\begin{equation}
\begin{split}
\hat{u}_a = B \hat{b}_{in}^{\dagger} + A_{a} \hat{a}_{in} - C_{a} \hat{m}_{in}^{\dagger},
\end{split}
\end{equation}
\begin{equation}
\begin{split}
\hat{u}_b = A_{b} \hat{b}_{in} - B \hat{a}_{in}^{\dagger} - C_{b} \hat{m}_{in},
\end{split}
\end{equation}
in which $A_j$, $B$, and $C_j$ $(j = a, b)$ are determined by opto- and electromagnonical cooperativity $\Lambda_a = G_{ma}^2 / \kappa_a \kappa_m$ and $\Lambda_b = g_{mb}^2 / \kappa_b \kappa_m$ as explicitly expressed in \citep{ref34}. $\hat{a}_{in}$, $\hat{b}_{in}$, and $\hat{m}_{in}$ are independent thermal states with their equilibrium mean thermal photon numbers $\bar{n}_a^T$, $\bar{n}_b^T$, and $\bar{n}_m^T$, respectively, governed by Planck's law. Due to the zero-mean Gaussian nature of the propagating modes $\hat{u}_a$ and $\hat{u}_b$, their photon number $\bar{n}_a$ and $\bar{n}_b$ and their cross-correlation read

$\bar{n}_a \equiv \left \langle \hat{u}_a^{\dagger}\hat{u}_a \right \rangle = \left | B \right |^2 (\bar{n}_b^T + 1) + \left | A_a \right |^2 \bar{n}_a^T + \left | C_a \right |^2 (\bar{n}_m^T + 1),$

$\bar{n}_b \equiv \left \langle \hat{u}_b^{\dagger}\hat{u}_b \right \rangle = \left | A_b \right |^2 \bar{n}_b^T + \left | B \right |^2 (\bar{n}_a^T + 1) + \left | C_b \right |^2 \bar{n}_m^T,$

$\left \langle \hat{u}_b\hat{u}_a \right \rangle = A_b B (\bar{n}_b^T + 1) - B A_a \bar{n}_a^T + C_a C_b (\bar{n}_m^T + 1)$.

The necessary and sufficient condition of the entanglement between propagating microwave and optical fields is that metric $\varepsilon \equiv \left | \left \langle \hat{u}_b\hat{u}_a \right \rangle \right | / \sqrt{\bar{n}_b \bar{n}_a}$ is bigger than one \citep{ref21, ref34}. As shown in Fig.2(a), one may find that the metric $\varepsilon$ satisfies this condition in the whole plane of the experimentally reachable cooperativity $\Lambda_a$ and $\Lambda_b$ with large values, suggesting the significant existence of the output signal-idler entanglement. The stability of the system is governed by the Routh-Hurwitz criterion \citep{ref34}, and all simulation parameters in this work satisfy this criterion.

To identify the quality of our microwave-optical entanglement source, only the existence of the propagating fields' entanglement is not enough. Before the quantification, one thing should be emphasized that the advantage of the QI strategy is computed on the basis of a fixed microwave photon number emitted toward the potential targets, since QI is an energy-restricted protocol where the quantum and classical strategies are compared with equal transmitted energy \citep{ref16, ref17}. Therefore, better quantum sources are expected to possess more quantum resources per microwave photon emitted \citep{ref21}. Under this condition, we shall use the normalized logarithmic negativity $ E_N / \bar{n}_b $ \citep{ref36, ref37, ref38} and normalized quantum coherent information $ I(a|b) / \bar{n}_b $ \citep{ref39} to quantify our magnon-based source, as shown in Fig.2(b) and Fig.2(c). These two quantify an upper and a lower bound of the average number of distillable entanglement bits for each microwave photon output from the source, respectively. Moreover, the normalized quantum discord $ D(b|a) / \bar{n}_b $, which catches the quantum correlations beyond the entanglement \citep{ref3, ref40}, are represented in Fig.2(d).

Compared with the EOM QI source \citep{ref21}, our magnonical QI strategy shows great outperformance on the quantum resources, as one may find from Fig.2 and Ref.\citep{ref21}. We attribute this mainly to the smaller average microwave photon number $\bar{n}_b$ per emitted mode. This is primarily caused by the enhanced resonant frequency of the medium that generates the microwave-optical entanglement, i.e., the magnon's resonant frequency is three orders of magnitude higher than the mechanical resonator in Ref.\citep{ref21}. Intuitively, the reason why a higher-frequency medium would cause fewer microwave photons per mode emitted is that, in the environment with identical temperature, the thermal excitation number of the higher-frequency magnon is smaller, leading to fewer quanta of the medium interacting with and generating the microwave-signal photons and optical-idler photons. Thus, with fewer generated microwave photons per mode, more normalized quantum resources compared with the EOM strategy could be obtained.

\textit{Quantum illumination and its phase-conjugate receiver.—}In the process of QI target detection depicted in Fig.1(b), $M = TW \gg 1$ independent and identically distributed (iid) microwave signal modes output from the continuous-wave magnonical converter with bandwidth $W$ (the common bandwidth of the entangled optical and microwave output fields is $W = \kappa_m(\Lambda_b-\Lambda_a+1)$ \citep{ref21}) are sent toward the spatial region of interest. Through the other converter, one may receive and collect the reflected $M$ microwave-signal mode to produce the converted optical field, and together with the corresponding $M$ retained idler to judge the presence (hypothesis $H_1$) or absence (hypothesis $H_0$) of the target by a $T$-sec-duration measurement. Under hypothesis $H_0$, the returned mode $\hat{u}_R$ of propagating microwave signal $\hat{u}_b$ will be $\hat{u}_R = \hat{u}_T$, where $u_T$ is in a thermal state with average photon number $\bar{n}_T \gg 1$. Under hypothesis $H_1$, the annihilation operator of the return-mode is $\hat{u}_R = \sqrt{\eta} \hat{u}_b + \sqrt{1-\eta} \hat{u}_T$, in which $0 < \eta \ll 1$ is the whole round-trip transmissivity and the thermal average photon number in this case is $\hat{n}_T / (1 - \eta) \approx \hat{n}_T$ \citep{ref16, ref21}. The expression of the $M$ iid converted optical modes read $\hat{u}_{\eta,a} = B \hat{u}_R^{\dagger} + A_a \hat{a}_{in}^{\prime} - C_a \hat{m}_{in}^{\prime \dagger}$, in which $\hat{a}_{in}^{\prime}$ and $\hat{m}_{in}^{\prime \dagger}$ possess identical thermal state with their counterparts in the former magnonical converter. Therefore, one can find that the receiver phase conjugates the reflected microwave field and converts it into the optical field. Then, the converted optical field together with the retained idler are input into a balanced beam splitter whose outputs are photodetected and fed into a unity-gain difference amplifier, producing an outcome to decide the presence or absence of the target with minimum error probability \citep{ref40}. For $M \gg 1$, the error probability is \citep{ref21, ref34, ref40} $P_{\textup{QI}}^{\textup{M}} = $ erfc$(\sqrt{\textup{SNR}_{\textup{QI}}^{\textup{M}}/8})/2$, where $\textup{SNR}_{\textup{QI}}^{\textup{M}}$ is the signal-to-noise ratio of the system for $M$-mode QI and erfc$(...)$ is the complementary error function.

To verify that QI's advantage still exists even though the microwave-optical entanglement breaks during the round trip, one needs to judge the existence of the entanglement between returned-microwave $\hat{u}_R$ and retained-optical $\hat{u}_a$ radiation. Under $H_0$, these two radiations will not get entangled, while under $H_1$ $\hat{u}_R$ and $\hat{u}_a$ will not get entangled when \citep{ref21} $\hat{n}_T \geq \hat{n}_T^{sill} \equiv (\left | \left \langle \hat{u}_b \hat{u}_a \right \rangle \right |^2 / \hat{n}_a - \hat{n}_b)$.

\begin{figure}
\centering
	\includegraphics[width=1\linewidth]{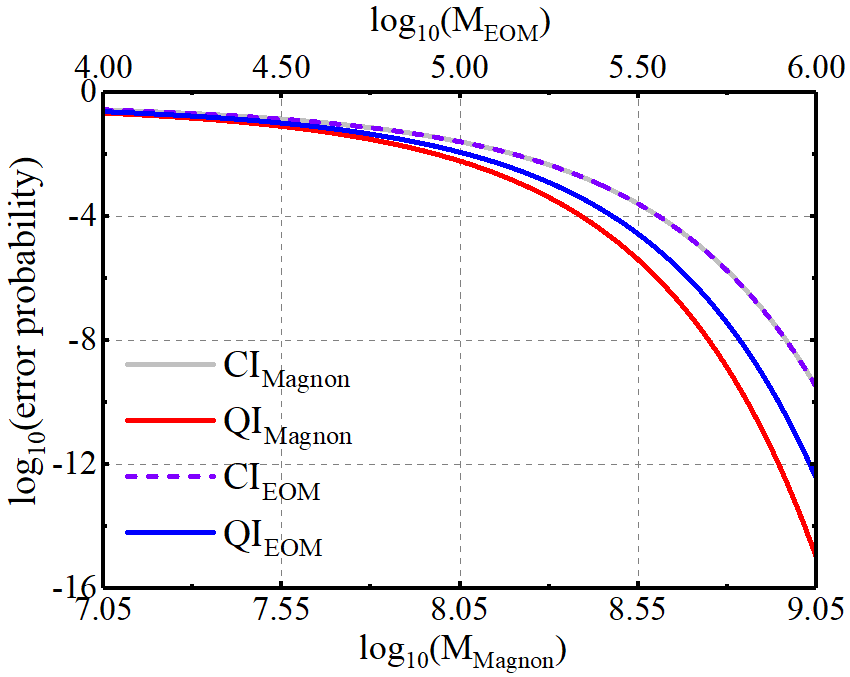}
	\caption{Error probability versus mode number or time-bandwidth product of the EOM $\textup{QI}_{\textup{EOM}}$ and its corresponding microwave transmitter $\textup{CI}_{\textup{EOM}}$, magnonical $\textup{QI}_{\textup{Magnon}}$ and its corresponding microwave transmitter $\textup{CI}_{\textup{Magnon}}$. $\textup{M}_{\textup{Magnon}}$ and $\textup{M}_{\textup{EOM}}$ are emitted microwave mode numbers for magnonical and EOM QI. For EOM: microwave cooperativity 5181.95 and optical cooperativity 668.43; for magnon-based scheme: microwave cooperativity 400 and optical cooperativity 0.054. The optical pumping wavelength is 1550 nm and the microwave resonant frequency is 9 GHz for both two quantum schemes. Assuming $\eta = $ 0.07 and room temperature $T_{\textup{room}} = $ 293 K (satisfying $\hat{n}_T \geq \hat{n}_T^{sill}$, which indicates that the returned microwave signal shares no entanglement with the retained optical idler), other parameters are the same as Fig.2.}
	\label{fig:Fig3.png}
\end{figure}

\textit{Comparison with EOM and classical microwave transmitters.—}Suppose three types of the transmitter (magnon-based, EOM and classical coherent-state microwave transmitter) all send $M\bar{n}_b$ photons on average to determine the target's existence or not, with $\bar{n}_b$ the mean microwave photon number per emitted mode. The minimum error probability of two quantum transmitter is $P_{\textup{QI}}^{\textup{M}}$ via the phase-conjugate receiver. The error probability of the classical microwave transmitter is \citep{ref41} $P_{\textup{CI}}^{\textup{M}} = $ erfc$(\sqrt{\textup{SNR}_{\textup{CI}}^{\textup{M}}/8})/2$ with $\textup{SNR}_{\textup{CI}}^{\textup{M}} = 4 \eta M \bar{n}_b / (2\bar{n}_T + 1)$ by homodyne detecting of the returned microwave field, which is the asymptotically the optimal receiver for the coherent-state transmitter for target detection \citep{ref21}.

The relationship between detection error probability and time-bandwidth product for three types of transmitters is shown in Fig.3. Due to the mean microwave-photon-number difference per mode between the EOM and magnon-based strategy, we plot the error-probability curve of their corresponding microwave transmitter with identical energy. Then, find appropriate $M$ for those two transmitters so that their corresponding microwave transmitter's performance to be the same, resulting in identical emitted energy between magnon-based and EOM transmitters. Under this condition, we could compare the performance between our magnon-based and EOM strategy. We find that, let alone two quantum strategies' great advantage over the classical one, the magnonical QI can possess orders of magnitude lower error probability compared with the EOM QI. In addition, as the EOM characteristic parameters change in a certain range, one could always find a counter magnonical parameter regime to obtain lower error probability, owing to more quantum resources per emitted microwave photon generated via magnonical system.

\begin{figure}
	\centering
	\includegraphics[width=1\linewidth]{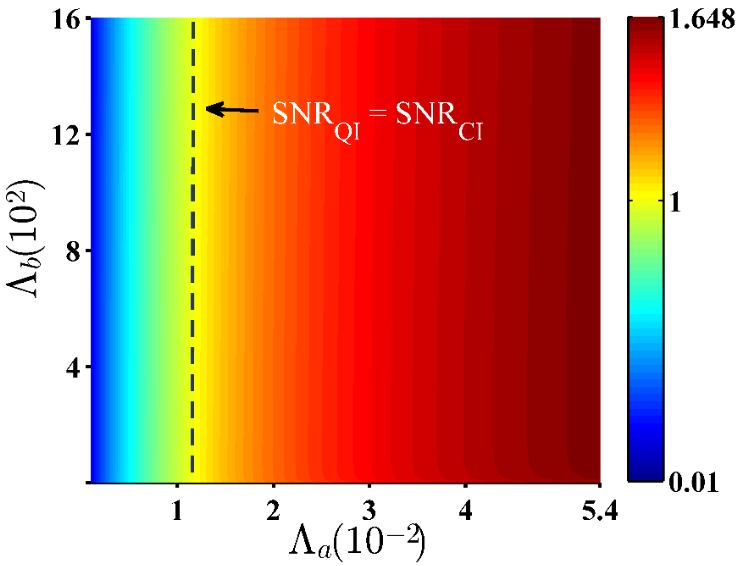}
	\caption{Magnonical QI's superiority $\mathcal{R}$ versus $\Lambda_a$ and $\Lambda_b$. For $\mathcal{R}$, the magnonical QI system possesses lower error probability than any classical system with equal transmitted energy. The black dashed line represents $\mathcal{R} = 1$, i.e., equal performance between QI and classical illumination. The other parameters are the same as Fig.2.}
	\label{fig:Fig4.png}
\end{figure}
To further analyze the advantage of our magnonical QI system over classical coherent-state microwave radar, we show the ratio $\mathcal{R} \equiv \textup{SNR}_{\textup{QI}}^{M} / \textup{SNR}_{\textup{CI}}^{M}$ for $M \gg 1$ in Fig.4. We notice that the quantum-enhanced sensitivity mainly depends on the optical cooperativity $\Lambda_a$ and exists when $\Lambda_a > 0.115$. This value coincides with the bound of $\Lambda_a$ that satisfies $G_{ma}^{2}/\kappa_a > \zeta$ (equivalent to $\Lambda_a > \bar{n}_b^T$), which guarantees the intra-cavity entanglement can be transmitted to the propagating microwave and optical fields. Once this condition is fulfilled, QI's advantage could emerge, and within an experimentally accessible parameter regime, larger $\Lambda_a$ would lead to better performance. Moreover, coherent-state system has the lowest error probability compared with other classical strategies containing equal emitted energy, indicating the magnonical QI's superiority over any classical schemes.

\textit{Conclusion and Discussion.—}We have shown that cavity magnonics, a promising platform studying quantum optics in solid-state systems and quantum information processing \citep{ref46, ref47, ref48}, can perform microwave QI. The magnonical converter discussed can generate significant amount of quantum entanglement, coherent information and quantum discord per emitted microwave photon between output microwave and optical fields. Despite the entanglement-breaking channel, such quantum resources could still lead to orders-of-magnitude lower detecting error probability compared with EOM QI and any classical-state microwave radar containing the same transmitted energy. The main difference between our scheme and the EOM QI scheme is the resonant frequency of the medium. The resonant frequency of magnon is about three orders of magnitude higher than the mechanical resonator used in Ref.\citep{ref21}, resulting in a smaller number of emitted microwave photon per mode that leads to more quantum resources per emitted microwave photon, which contributes to lower detecting error probability with equal emitted energy. Based on this, we may conclude that other medium with a high resonant frequency that is able to perform microwave-optical quantum transduction can also provide this advantage on QI, like optomechanical nanobeam \citep{ref42}.

Note that the emitted microwave mode number $M = TW$, which means, with fixed common bandwidth $W$ of the entangled optical and microwave fields, if we need a larger $M$ to obtain better quantum-enhanced sensitivity, the longer measurement time $T$ will be taken. To reduce the measurement time while keeping the identical detecting error probability, we can resort to enhance the optical cooperativity $\Lambda_a = G_{ma}^2 / \kappa_a \kappa_m$. The potential solutions are that utilize other magnetic materials with larger Verdet constant, like $\chemfig{CrBr_3}$ with Verdet constant 8700 rads/cm for the light at 500 nm in 1.5 K environment, to replace the YIG for enhancing the optomagnonical coupling $G_{ma}$, or turn to advanced micro- and nanofabrication technology to reduce the surface roughness, thus enhancing the quality factor $Q$ of YIG sphere for WGM. Moreover, the integrated cavity optomagnonical waveguide also shows its potential to realize high magneto-optical cooperativity \citep{ref43}. 

The receiver used in this work is the phase-conjugate receiver which limits our detecting range within 11.25 km in free space (assuming fiber loss 0.2 dB/km, and optical fields' propagation speed in fiber 2$c$/3, where $c$ is vacuum light speed) \citep{ref21}, since the retained idler loss cannot exceed 3 dB, otherwise the quantum advantage of target detection would disappear \citep{ref41}. Though the detecting range is limited, our scheme could still be useful in quantum-enhanced catching the objects in the near region for security or in the biological and medical applications. Moreover, there is a receiver design that could acheive predicted optimum 6 dB advantage of QI over optimum classical sensing \citep{ref49} and has successfully shown its advantage in certain scenarios \citep{ref50, ref51}, which is also a promising road to enhance the microwave QI together with our proposal. With initial microwave QI experiments emerge \citep{ref22, ref44, ref45}, exploring which type of receivers can be implemented in the microwave domain is still an open question and may spur continued research on quantum radar.
\\

This work has been supported by National Key R$\textup{\&}$D Program of China (2018YFA0307400); National Natural Science Foundation of China (NSFC) (61775025, U19A2076).

\bibliography{Microwave.bib}
\end{document}


\title{Supplimental Material for ``Microwave Quantum Illumination via Cavity Magnonics"}
\author{Qizhi Cai$^{1,2}$}
\author{Jinkun Liao$^{1,2}$}\thanks{E-mail: jkliao@uestc.edu.cn}
\author{Bohai Shen$^{1}$}
\author{Guangcan Guo$^{1,3}$}
\author{Qiang Zhou$^{1,2,3,}$}\thanks{E-mail: zhouqiang@uestc.edu.cn}
\address{$^{1}$Institute of Fundamental and Frontier Sciences, University of Electronic Science and Technology of China, Chengdu, Sichuan, China}
\address{$^{2}$School of optoelectronic science and engineering, University of Electronic Science and Technology of China, Chengdu, Sichuan, China}
\address{$^{3}$CAS Key Laboratory of Quantum Information, University of Science and Technology of China, Hefei, China}

\date{\today}
\begin{abstract}
This Supplemental Material contains the technical details, including (i) description of the interaction Hamiltonian of a magnon-based system and its related parameters, (ii) quantum Langevin equations and stability of the system, (iii) quantum resources of our magnon-based QI source, (iv) derivation of the error probability for $M$ microwave-optical mode pairs.
\end{abstract}

\pacs{Valid PACS appear here}
\maketitle

\setcounter{equation}{0}
\setcounter{figure}{0}
\setcounter{table}{0}
\setcounter{page}{1}
\makeatletter
\renewcommand{\theequation}{S\arabic{equation}}
\renewcommand{\thefigure}{S\arabic{figure}}
\renewcommand{\bibnumfmt}[1]{[S#1]}
\renewcommand{\citenumfont}[1]{S#1}

\section{I. Description of the interaction Hamiltonian and related parameters}
Our magnon-based microwave-optical hybrid system consists of a yttrium iron garnet (YIG) sphere that holds the magnon modes; a microwave cavity in which the resonant microwave cavity mode can interact with the magnon mode via magnetic dipole interaction, building the microwave-magnon coupling; a nanofiber where the pump optical field inside can evanescently couple to the whispering gallery mode (WGM) in the YIG sphere, the WGM shares a magnon-induced Brillouin scattering process with the magnon mode, establishing the optomagnonical coupling \citep{S_ref1, S_ref2, S_ref3, S_ref4, S_ref5, S_ref6, S_ref7}.

For the optomagnonical part, the situation we consider is that the WGM photons interact with the magnon modes as the external magnetic field is applied perpendicularly to the WGM orbit plane. In this way, magnon modes could interact with almost purely $\sigma^{+}$-, $\sigma^{-}$- or $\pi$-polarized optical photon depending on the polarization of optical photons and the direction of the WGM orbit. The direction of WGM orbits has two: counterclockwise (CCW) orbit corresponding to (TE, TM) = ($\pi$, $\sigma^{+}$) resonant in the YIG sphere and clockwise (CW) orbit corresponding to (TE, TM) = ($\pi$, $\sigma^{-}$). Assuming the annihilation operators of the TE, TM WGM, and magnon are $\hat{a}^{ }_{\textup{TE}}$, $\hat{a}^{ }_{\textup{TM}}$ and $\hat{m}$. In the CCW case, the interaction Hamiltonian reads
\begin{equation}
\begin{split}
H^{CCW}_{int} = \hbar g_{ma} (\hat{a}^{\dagger}_{\textup{TM}} {\hat{a}^{ }}_{\textup{TE}} \hat{m} + {\hat{a}^{ }}_{\textup{TM}} \hat{a}^{\dagger}_{\textup{TE}} \hat{m}^{\dagger}),
\end{split}
\end{equation}
with coupling constant $g_{ma}$, which means that if the input photon is in the TM mode with angular frequency $\omega$ and $\sigma^{+}$-polarization, via the magnon-based Brillouin scattering, one magnon with angular frequency $\omega_{mag}$ and one down-converted photon with $\pi$-polarization in the TE mode with angular frequency $\omega-\omega_{mag}$ are generated, fulfilling the conservation of energy and spin angular momentum. In the CW case that is symmetrical with the CCW one, the interaction Hamiltonian is
\begin{equation}
\begin{split}
H^{CW}_{int} = \hbar g_{ma} (\hat{a}^{\dagger}_{\textup{TE}} {\hat{a}^{ }}_{\textup{TM}} \hat{m} + {\hat{a}^{ }}_{\textup{TE}} \hat{a}^{\dagger}_{\textup{TM}} \hat{m}^{\dagger}),
\end{split}
\end{equation}
and the interaction picture reads that if the input photon is in the TE mode with $\pi$-polarization, via the magnon-based Brillouin scattering, one magnon and one down-converted photon with $\sigma^{-}$-polarization in the TM mode would be created \citep{S_ref1}. For simplicity and without loss of generality, we only choose the CW case in the main text, which means we would only pump the optical field that couples the TE WGM in the YIG sphere.

The optomagnonical coupling rate is \citep{S_ref1}
\begin{equation}
\begin{split}
{g_{ma}} = \mathcal{V}\frac{c}{{{n_r}}}\sqrt {\frac{2}{{{n_{spin}}V_{sp}}}},
\end{split}
\end{equation}
where the YIG's Verdet constant $\mathcal{V}$ = 3.77 rad/cm, refractive index $n_r$ = 2.19, and spin density $n_{spin}$ = 2.1$\times {10^{28}}/{m^3}$. $V_{sp}=\frac{4\pi}{3}r^{3}$ is the volume of the YIG sphere, $r$ is the radius of YIG sphere, and $c$ is the speed of light in vacuum. Supposing that the TE mode ($\hat{a}^{ }_{\textup{TE}}$) is driving with a bright coherent tone, the interaction Hamiltonian in CW case becomes $H^{CW}_{int} = \hbar G_{ma} ({\hat{a}^{ }}_{\textup{TM}} \hat{m} + \hat{a}^{\dagger}_{\textup{TM}} \hat{m}^{\dagger})$ with ${\hat{a}^{ }}_{\textup{TE}} \rightarrow \alpha$ and $G_{ma} \equiv g_{ma} \alpha$. The intra-cavity photon number of TE WGM is $N_{\textup{TE}} = \alpha^2 = \frac{2}{\kappa_a} \frac{P_p}{\hbar \omega_p}$ with optical-TE damping rate $\kappa_a$, and pump angular frequency $\omega_p$, where $P_p$ = 60 mW is the optical-TE pump power, pump wavelength $\lambda_p = \frac{2 \pi c}{\omega_p}$ = 1550 nm, WGM quality factor $Q = \omega_{\textup{TE}} / \kappa_a = 3 \times 10^6$. With the radius of the YIG sphere $r =$ 100 $\mu$m and damping rate of magnon $\kappa_m / 2\pi$ = 1 MHz, the corresponding cooperativity of TM WGM $\Lambda_a = G_{ma}^2 / \kappa_a \kappa_m \approx $ 0.055 (the maximum value of $\Lambda_a$ in Fig.2 of the main text is 0.054). As shown in the Fig.4 in the main text, larger $\Lambda_a$ in a certain range would advance the performance of the QI. To enlarge the value of $\Lambda_a$, one can resort to increase the pump power $P_p$, increase the quality factor $Q$, and reduce the radius $r$ of the YIG sphere from the theoretical perspective.

For the electromagnonical part, the magnon-microwave interaction naturally occurs on beam-splitter form in experiment \citep{S_ref2, S_ref3, S_ref4, S_ref5} $\hbar g_{mb}(\hat{b}\hat{m}^{\dagger}+\hat{b}^{\dagger}\hat{m})$ with the annihilation operator of microwave mode $\hat{b}$, and the electromagnonical coupling can reach a strong-coupling regime. We choose $g_{mb}$ = 40 MHz and get the corresponding eletromagnonical cooperativity $\Lambda_b = g_{mb}^2 / \kappa_b \kappa_m$ = 1600 with the microwave damping rate $\kappa_b$= 1 MHz. The common bandwidth of entangled microwave and optical fields in our magnon-based converter is $W = \kappa_m(\Lambda_b - \Lambda_a + 1)$ \citep{S_ref8}. Due to $\Lambda_b \gg \Lambda_a$, the dominant factor of bandwidth $W$ is the microwave cooperativity, and larger $\Lambda_b$ would result in larger $W$, bringing practical advantages of our converter in the goal of QI target detection.

\section{II. Quantum Langevin equations and Routh-Hurwitz criterion of the system}
As shown in the main text, the linearized interaction Hamiltonian is
\begin{equation}
\begin{split}
H^{'} = \hbar G_{ma} (\hat{a}^{\dagger} \hat{m}^{\dagger} + \hat{a} \hat{m}) +\hbar g_{mb}(\hat{b} \hat{m}^{\dagger}+\hat{b} ^{\dagger}\hat{m}),
\end{split}
\end{equation}
in which we have denoted ${\hat{a}^{ }}_{\textup{TM}}$ as $\hat{a}$ for simplicity. The total quantum treatment of the system can be described by the quantum Langevin equations where we add the Heisenberg equations with the quantum noise acting on magnon mode ($\hat{m}_{in}$ with damping rates $\kappa_m$) as well as input noises of optical and microwave mode ($\hat{j}_{in}$ with damping rates $\kappa_j$, $j = a, b$ represents the optical and microwave mode respectively). These noises operators are zero mean and characterized by the following correlation functions
\begin{equation}
\begin{split}
&\left \langle \hat{a}_{in}(t) \hat{a}_{in}^{\dagger}(t')\right \rangle = \left \langle \hat{a}_{in}^{\dagger}(t) \hat{a}_{in}(t')\right \rangle + \delta(t-t') = (\bar{n}_a^T+1)\delta(t-t'),\\
&\left \langle \hat{b}_{in}(t) \hat{b}_{in}^{\dagger}(t')\right \rangle = \left \langle \hat{b}_{in}^{\dagger}(t) \hat{b}_{in}(t')\right \rangle + \delta(t-t') = (\bar{n}_b^T+1)\delta(t-t'),\\
&\left \langle \hat{m}_{in}(t) \hat{m}_{in}^{\dagger}(t')\right \rangle = \left \langle \hat{m}_{in}^{\dagger}(t) \hat{m}_{in}(t')\right \rangle + \delta(t-t') = (\bar{n}_m^T+1)\delta(t-t'),
\end{split}
\end{equation}
where $\bar{n}_a$, $\bar{n}_b$, and $\bar{n}_m$ are the thermal excited numbers of each bath. The resulting quantum Langevin equations for the intra-cavity modes and magnon mode are
\begin{equation}
\begin{split}
&\dot{\hat{a}} = - \kappa_a \hat{a} - i G_{ma} \hat{m}^{\dagger} + \sqrt {2{\kappa_a}} {\hat{a}_{in}},\\
&\dot{\hat{b}} = - \kappa_b \hat{b} - i g_{mb} \hat{m} + \sqrt {2{\kappa _b}} {\hat{b}_{in}},\\
&\dot{\hat{m}} = - \kappa_m \hat{m} - i G_{ma} \hat{a}^{\dagger} - i g_{mb} \hat{b} + \sqrt {2{\kappa _m}} {\hat{m}_{in}}.
\end{split}
\end{equation}
In this paper, we are interested in the entanglement of output microwave and optical fields, which can be calculated by utilizing the input-output theory. Therefore for simplicity, we solve the Eq.(S6) by moving to the frequency domain, obtaining the microwave and optical cavities' variables. Then, substituting the solutions of Eq.(S6) into the input-output formula, i.e., $\hat{u}_j \equiv \hat{j}_{\textup{out}} = \sqrt{2\kappa_j} \hat{j} - \hat{j}_{in}$, we get
\begin{equation}
\begin{split}
&\hat{u}_a = B(\omega) \hat{b}_{in}^{\dagger} + A_{a}(\omega) \hat{a}_{in} - C_{a}(\omega) \hat{m}_{in}^{\dagger},\\
&\hat{u}_b = A_{b}(\omega) \hat{b}_{in} - B(\omega) \hat{a}_{in}^{\dagger} - C_{b}(\omega) \hat{m}_{in},
\end{split}
\end{equation}
in which 
\begin{equation}
\begin{split}
&A_a(\omega) = \frac{-[\widetilde{\omega}_a^{*}-2][\Lambda_b+\widetilde{\omega}_b^{*}\widetilde{\omega}_m^{*}]+\Lambda_a\widetilde{\omega}_b^{*}}{\widetilde{\omega}_b^{*}[\widetilde{\omega}_a^{*}\widetilde{\omega}_m^{*}-\Lambda_a]+\Lambda_b\widetilde{\omega}_a^{*}},\\
&A_b(\omega) = \frac{[\widetilde{\omega}_b-2][\Lambda_a-\widetilde{\omega}_a\widetilde{\omega}_m]-\Lambda_b\widetilde{\omega}_a}{\widetilde{\omega}_b[\widetilde{\omega}_a \widetilde{\omega}_m-\Lambda_a]+\Lambda_b\widetilde{\omega}_a},\\
&B(\omega) = \frac{2\sqrt{\Lambda_a \Lambda_b}}{\widetilde{\omega}_b[\widetilde{\omega}_a\widetilde{\omega}_m-\Lambda_a]+\Lambda_b\widetilde{\omega}_a},\\
&C_a(\omega) = \frac{2i\sqrt{\Lambda_a}\widetilde{\omega}_b^{*}}{\widetilde{\omega}_b^{*}[\widetilde{\omega}_a^{*}\widetilde{\omega}_m^{*}-\Lambda_a]+\Lambda_b\widetilde{\omega}_a^{*}},\\
&C_b(\omega) = \frac{2i\sqrt{\Lambda_b}\widetilde{\omega}_a}{\widetilde{\omega}_b[\widetilde{\omega}_a\widetilde{\omega}_m-\Lambda_a]+\Lambda_b\widetilde{\omega}_a},
\end{split}
\end{equation}
with $\widetilde{\omega}_j = 1 - i \omega / \kappa_j$ and $\widetilde{\omega}_m = 1 - i \omega / \kappa_m$ (we have assumed that the internal losses are negligible). The coefficients Eq.(S8) would be simpler when $\omega \approx 0$, corresponding to take a narrow frequency band around each cavity resonance, namely
\begin{equation}
\begin{split}
&A_a = \frac{1+(\Lambda_a+\Lambda_b)}{1+\Lambda_b-\Lambda_a},\\
&A_b = \frac{1-(\Lambda_a+\Lambda_b)}{1+\Lambda_b-\Lambda_a},\\
&B = \frac{2\sqrt{\Lambda_a\Lambda_b}}{1+\Lambda_b-\Lambda_a},\\
&C_a = \frac{2i\sqrt{\Lambda_a}}{1+\Lambda_b-\Lambda_a},\\
&C_b = \frac{2i\sqrt{\Lambda_b}}{1+\Lambda_b-\Lambda_a},
\end{split}
\end{equation}
and Eq.(S7) reduces to the expression presented in the main text
\begin{equation}
\begin{split}
&\hat{u}_a = B \hat{b}_{in}^{\dagger} + A_{a} \hat{a}_{in} - C_{a} \hat{m}_{in}^{\dagger},\\
&\hat{u}_b = A_{b} \hat{b}_{in} - B \hat{a}_{in}^{\dagger} - C_{b} \hat{m}_{in}.
\end{split}
\end{equation}
The input-output operations preserve the bosonic commutations between the operators, i.e., $[\hat{u}_k,\hat{u}_l^{\dagger}] = \delta_{k,l}$ and $[\hat{u}_k,\hat{u}_l] = [\hat{u}_k^{\dagger},\hat{u}_l^{\dagger}] = 0$ for $(k,l \in a, b)$.

The linearized Heisenberg-Langevin equations for our intra-cavity and magnon modes are given in Eq.(S6). For simple calculation of the stability of the system, we introduce the quadrature for each mode in Eq.(S6): $\hat X_m = ( \hat{m} + \hat{m}^{\dagger} ) / \sqrt{2}$ and $\hat Y_m = ( \hat{m} - \hat{m}^{\dagger} ) / i \sqrt{2}$ for the magnon mode, $\hat X_a = ( \hat{a} + \hat{a}^{\dagger} ) / \sqrt{2}$ and $\hat Y_a = ( \hat{a} - \hat{a}^{\dagger} ) / i \sqrt{2}$ for the optical mode, and $\hat X_b = ( \hat{b} + \hat{b}^{\dagger} ) / \sqrt{2}$ and $\hat Y_b = ( \hat{b} - \hat{b}^{\dagger} ) / i \sqrt{2}$ for the microwave mode. After dropping the noise terms, Eq.(S6) becomes $\dot{\vec{f}}$ = $\textup{M} \cdot \vec{f}$ in which $\vec{f} = (\hat{X}_a, \hat{Y}_a, \hat{X}_b, \hat{Y}_b, \hat{X}_m, \hat{Y}_m)$ and 
\begin{equation}
\begin{split}
\textup{M}=\begin{pmatrix}
-\kappa_m & 0 & 0 & -G_{ma} & 0 & g_{mb}\\ 
0 & -\kappa_m & -G_{ma} & 0 & -g_{mb} & 0\\ 
0 & -G_{ma} & -\kappa_a & 0 & 0 & 0\\ 
-G_{ma} & 0 & 0 & -\kappa_a & 0 & 0\\ 
0 & g_{mb} & 0 & 0 & -\kappa_b & 0\\ 
-g_{mb} & 0 & 0 & 0 & 0 & -\kappa_b
\end{pmatrix}.
\end{split}
\end{equation}
The stability of the system is checked by the Routh-Hurwitz criterion \citep{S_ref9, S_ref10}. That is, if all eigenvalues of $\textup{M}$ hold negative real parts, the system will reach a steady state. However, the exact expression is too cumbersome, so we omit it here. All parameters used in this work will fulfill the Routh-Hurwitz criterion, which indicates that the system will always be stable in numerical simulation.

\section{III. Microwave-optical quantum resources}
\subsection{A. Output-fields Entanglement Metric}
From Eq.(S10), we can know that the output fields $\hat{u}_a$ and $\hat{u}_b$ are determined by a collection of independent and thermal-state inputs $\hat{a}_{in}$, $\hat{b}_{in}$ and $\hat{m}_{in}$. Therefore, $\hat{u}_a$ and $\hat{u}_b$ are in a zero-mean, jointly Gaussian state that is totally characterized by the non-zero second moments
\begin{equation}
\begin{split}
&\bar{n}_a \equiv \left \langle \hat{u}_a^{\dagger}\hat{u}_a \right \rangle = \left | B \right |^2 (\bar{n}_b^T + 1) + \left | A_a \right |^2 \bar{n}_a^T + \left | C_a \right |^2 (\bar{n}_m^T + 1),\\
&\bar{n}_b \equiv \left \langle \hat{u}_b^{\dagger}\hat{u}_b \right \rangle = \left | A_b \right |^2 \bar{n}_b^T + \left | B \right |^2 (\bar{n}_a^T + 1) + \left | C_b \right |^2 \bar{n}_m^T,\\
&\left \langle \hat{u}_b\hat{u}_a \right \rangle = A_b B (\bar{n}_b^T + 1) - B A_a \bar{n}_a^T + C_a C_b (\bar{n}_m^T + 1).
\end{split}
\end{equation}

The sufficient and necessary condition of that the joint Gaussian state will be classical is the phase-sensitive cross correlation $\left | \left \langle \hat{u}_b\hat{u}_a \right \rangle \right |$ satisfies the classical bound \citep{S_ref8}
\begin{equation}
\begin{split}
\left | \left \langle \hat{u}_b\hat{u}_a \right \rangle \right | \leq \sqrt{\bar{n}_b \bar{n}_a}.
\end{split}
\end{equation}
If $\left | \left \langle \hat{u}_b\hat{u}_a \right \rangle \right |$ violates this bound, the output optical $\hat{u}_a$ and microwave $\hat{u}_b$ modes would get entangled. This classical bound derived from the inequality $\left \langle \left | z\hat{u}_a + \hat{u}_b^{*} \right |^2 \right \rangle \geq 0$ for classical and complex random variables $\hat{u}_a$ and $\hat{u}_b$, in which $\left \langle \cdots \right \rangle$ is the classical average and $z$ is a random real-valued parameter.

Figure 2(a) in the main text plot the relation of output-fields Entanglement Metric $\varepsilon \equiv \left | \left \langle \hat{u}_b\hat{u}_a \right \rangle \right | / \sqrt{\bar{n}_b \bar{n}_a}$ versus $\Lambda_a$ and $\Lambda_b$. $\varepsilon > 1$ indicates that the output microwave and optical modes are entangled, under experimentally reachable parameter setting, one may find that our magnon-based microwave-optical source possesses great entanglement properties ($\varepsilon \gg 1$) and satisfies Eq.(S12) in all region.

\subsection{B. Logarithmic Negativity}
Here we measure the entanglement between output optical and microwave fields by using the logarithmic negativity \citep{S_ref11, S_ref12}, which is the upper bound of the number of distillable entanglement bits generated by our source. Since quantum illumination (QI) is an energy-restricted protocol, where the quantum and classical strategies are compared with the identical mean photon number of the signal mode, we normalize the entanglement measure with the mean number of microwave photon $\bar{n}_b$ emitted \citep{S_ref8}. We expect that a better source will contain larger entanglement or other quantum resources per microwave photon emitted.

For computing the logarithmic negativity, we first determine the covariance matrix (CM) frequency domain, the elements read
\begin{equation}
\begin{split}
\delta(\omega + {\omega}') V_{ij}(\omega) = \frac{1}{2} \left \langle \Delta v_i(\omega) \Delta v_j({\omega}') + \Delta v_j({\omega}') \Delta v_i(\omega) \right \rangle,
\end{split}
\end{equation}
where $\Delta v_i = v_i - \left \langle v_i \right \rangle$, $v(\omega) = [X_{u_b}(\omega), Y_{u_b}(\omega), X_{u_a}(\omega), Y_{u_a}(\omega)]^{T}$, $X_{u_q} = (\hat{u}_q + \hat{u}_q^{\dagger}) / \sqrt{2}$, and $Y_{u_q} = (\hat{u}_q - \hat{u}_q^{\dagger}) / i\sqrt{2}$ with $(q = a, b)$ \citep{S_ref8, S_ref14, S_ref15}.

Then, by using Eq.(S10) and Eq.(S14), we obtain the CM for the quadratures of the optical and microwave output fields
\begin{equation}
\begin{split}
\mathbf{V}(\omega) = \begin{pmatrix}
V_{11} & 0 & V_{13} & 0\\ 
0 & V_{11} & 0 & -V_{13}\\ 
V_{13} & 0 & V_{33} & 0\\ 
0 & -V_{13} & 0 & V_{33}
\end{pmatrix},
\end{split}
\end{equation}
where
\begin{equation}
\begin{split}
&V_{11} = \frac{\left \langle X_{u_b}(\omega)X_{u_b}({\omega}') \right \rangle}{\delta(\omega+{\omega}')}=\bar{n}_b + 1/2,\\
&V_{33} = \frac{\left \langle X_{u_a}(\omega)X_{u_a}({\omega}') \right \rangle}{\delta(\omega+{\omega}')}=\bar{n}_a + 1/2,\\
&V_{13} = \frac{\left \langle X_{u_b}(\omega)X_{u_a}({\omega}') + X_{u_a}({\omega}')X_{u_b}(\omega) \right \rangle}{2\delta(\omega+{\omega}')} = \left \langle \hat{u}_b\hat{u}_a \right \rangle,
\end{split}
\end{equation}
where we have used the fact that $\left \langle \hat{u}_a^2 \right \rangle = \left \langle \hat{u}_b^2 \right \rangle = 0$ and $\left \langle \hat{u}_b\hat{u}_a \right \rangle$ is real-valued. One can easily find that Eq.(S15) is two-mode squeezed thermal state CM that can generate entanglement between the two modes it describes.

The logarithmic negativity $E_N$ is given by \citep{S_ref11, S_ref12}
\begin{equation}
\begin{split}
E_{N} = \max[0,-\ln 2 \xi^{-}],
\end{split}
\end{equation}
in which $\xi^{-}$ is the smallest partially transposed symplectic eigenvalue of $\mathbf{V}(\omega)$, expressed as \citep{S_ref8, S_ref14}
\begin{equation}
\begin{split}
\xi^{-} = 2^{-1/2}\left ( V_{11}^2 + V_{33}^2 + 2V_{13}^2 - \sqrt{(V_{11}^2 - V_{33}^2)^2 + 4V_{13}^2(V_{11}^2+V_{33}^2)^2} \right )^{1/2}.
\end{split}
\end{equation}
In Fig.2(b) of the main text, we have shown the normalized logarithmic negativity $E_N / \bar{n}_b$ versus the optical cooperativity $\Lambda_a$ and the microwave cooperativity $\Lambda_b$. We find that our source has a good quality in terms of the distillable entanglement bits per emitted microwave photon and keeps this advantage in all region of the panel.

\subsection{C. Coherent Information}
Now we compute the coherent information of our magnon-based microwave-optical source, which corresponds to the lower bound to the number of distillable entanglement bits generated by our source. This is given by \citep{S_ref16}
\begin{equation}
\begin{split}
I(a|b) = S(b) - S(a,b),
\end{split}
\end{equation}
where $S(b)$ is the von Neumann entropy of the microwave mode and $S(a,b)$ is the joint von Neumann entropy of the optical and microwave modes \citep{S_ref14}. Moreover, in the light of hasing inequality \citep{S_ref8, S_ref17}, $I(a|b)$ represents the distillable rate of entanglement bits per use of the source directed from the optical to the microwave part.

For our Gaussian source, we can simply give the entropy terms in Eq.(S19) with the symplectic eigenvalue $\nu_b = V_{11}$ of the reduced CM related to the microwave mode and the total CM $\mathbf{V}(\omega)$'s symplectic spectrum $\left \{ \nu_+, \nu_- \right \}$ \citep{S_ref14}
\begin{equation}
\begin{split}
\nu_{\pm } = 2^{-1/2}\left ( V_{11}^2 + V_{33}^2 - 2V_{13}^2 \pm \sqrt{(V_{11}^2 - V_{33}^2) - 4 V_{13}^2(V_{11}-V_{33})^2} \right )^{1/2}.
\end{split}
\end{equation}
Therefore, we have \citep{S_ref14}
\begin{equation}
\begin{split}
I(a|b) = h(V_{11}) - h(\nu_+) - h(\nu_-),
\end{split}
\end{equation}
in which (Note that our notation is different from the Ref.\citep{S_ref14} where the vacuum noise of quadratures $\Delta v_i$ is 1. In our notation, the corresponding vacuum noise is 1/2.)
\begin{equation}
\begin{split}
h(x) \equiv (x + \frac{1}{2}) \log_2 (x + \frac{1}{2}) - (x - \frac{1}{2}) \log_2 (x - \frac{1}{2}).
\end{split}
\end{equation}

In Fig.2(c) of the main text shows the normalized coherent information $I(a|b) / \bar{n}_b$ versus the optical cooperativity $\Lambda_a$ and the microwave cooperativity $\Lambda_b$. We find that our source has a good quality in terms of the distillable entanglement bits (lower bound) per emitted microwave photon and keeps this advantage in all region of the panel.

\subsection{D. Quantum Discord --- Quantum Correlations Beyond Entanglement}
Since the quantum state generated by our microwave-optical source is a mixed Gaussian state, as one can readily examine from the value of joint von Neumann entropy $S(a,b)$, it is necessary to determine the quality of our source in terms of general quantum correlations beyond quantum entanglement. Therefore, we compute the normalized quantum discord \citep{S_ref19, S_ref20} $D(b|a) / \bar{n}_b$ of the source, which captures the quantum correlations that are carried by each microwave photon emitted toward the potential targets.

According to Ref.\citep{S_ref10}, the CM in Eq.(S15) can be rewritten as
\begin{equation}
\begin{split}
\mathbf{V}(\omega)=\begin{pmatrix}
(\tau \tilde{b} + \tilde{\eta}) \mathbf{I}& \sqrt{\tau(\tilde{b}^2-1)}\mathbf{C}\\ 
\sqrt{\tau(\tilde{b}^2-1)}\mathbf{C} & \tilde{b} \mathbf{I}
\end{pmatrix},\mathbf{I} \equiv \textup{diag}(1,1),\mathbf{C} \equiv \textup{diag}(1,-1),
\end{split}
\end{equation}
where
\begin{equation}
\begin{split}
\tilde{b} = V_{33}, \tau = \frac{V_{13}^2}{V_{33}^2-1}, \tilde{\eta} = V_{11} - \frac{V_{33}V_{13}^2}{V_{33}^2-1}.
\end{split}
\end{equation}
Then, we have \citep{S_ref20}
\begin{equation}
\begin{split}
D(b|a) &= h(\tilde{b}) - h(\nu_+) - h(\nu_-) + h(\tau + \tilde{\eta})\\ 
&= h(V_{33}) - h(\nu_+) - h(\nu_-) + h\left ( V_{11} + \frac{V_{13}^2(1-V_{33})}{V_{33}^2-1} \right ),
\end{split}
\end{equation}
in which $\nu_+$ and $\nu_-$ are the symplectic eigenvalues of $\mathbf{V}(\omega)$ as shown in Eq.(S20).

In Fig.2(d) of the main text shows the normalized quantum discord $D(b|a) / \bar{n}_b$ versus the optical cooperativity $\Lambda_a$ and the microwave cooperativity $\Lambda_b$. We can see that our source has a good quality in terms of the quantum discord per emitted microwave photon and keeps this advantage in all region of the panel.

\section{IV. Error probability of the system for $M$ microwave-optical mode pairs}
The input-output relation of our magnon-based microwave-optical converter is given by Eq.(S10) with their coefficients as shown in Eq.(S9). Therefore, the QI receiver's input-output relation reads
\begin{equation}
\begin{split}
\hat{u}_{\eta,a} = B \hat{u}_R^{\dagger} + A_a \hat{a}_{in}^{\prime} - C_a \hat{m}_{in}^{\prime \dagger},
\end{split}
\end{equation}
in which $\hat{a}_{in}^{\prime}$ and $\hat{m}_{in}^{\prime \dagger}$ have the same thermal state with their counterparts given in Eq.(S5). The QI receiver collects $M$ independent and identically distributed microwave signal modes $\left \{ \hat{u}_R^{(k)} : 1 \leq k \leq M \right \}$ and converts them into optical domain $\left \{ \hat{u}_{\eta, a}^{(k)} \right \}$. As shown in Fig.1, the converted optical signal $\hat{u}_{\eta, a}^{(k)}$ and the retained $M$ optical idler $\left \{ \hat{u}_{a}^{(k)} \right \}$ are mixed by a 50-50 beam splitter whose outputs are
\begin{equation}
\begin{split}
\hat{u}_{\eta, \pm }^{(k)} \equiv \frac{\hat{u}_{\eta,a}^{(k)}\pm \hat{u}_a^{(k)}}{\sqrt{2}}.
\end{split}
\end{equation}
Note that $\left \{ \hat{u}_{\eta,a}^{(k)}, \hat{u}_{\eta,a}^{(k)} : 1 \leq k \leq M \right \}$ are in a zero-mean, jointly-Gaussian state that is characterized by second momoents $\left \langle \hat{u}_{a}^{(k)\dagger} \hat{u}_{a}^{(k)} \right \rangle$, $\left \langle \hat{u}_{\eta,a}^{(k)\dagger} \hat{u}_{\eta,a}^{(k)} \right \rangle_{H_j}$, and $\left \langle \hat{u}_{\eta,a}^{(k)} \hat{u}_{a}^{(k)} \right \rangle_{H_j}$, for $j=0,1$. Then, these modes are detected, yielding the photon counts that are equal to the measurements of the corresponding number operator $\hat{N}_{\eta, \pm }^{(k)} \equiv \hat{u}_{\eta, \pm }^{(k)\dagger} \hat{u}_{\eta, \pm }^{(k)}$. Finally, the decision of the target presense or absense is made by comparing the difference of the total photon counts of two detectors
\begin{equation}
\begin{split}
\hat{N}_{\eta}=\sum_{k=1}^{M} (\hat{N}_{\eta, + }^{(k)} - \hat{N}_{\eta, - }^{(k)}).
\end{split}
\end{equation}
\begin{figure}[t!]
	\centering
	\includegraphics[width=0.85\linewidth]{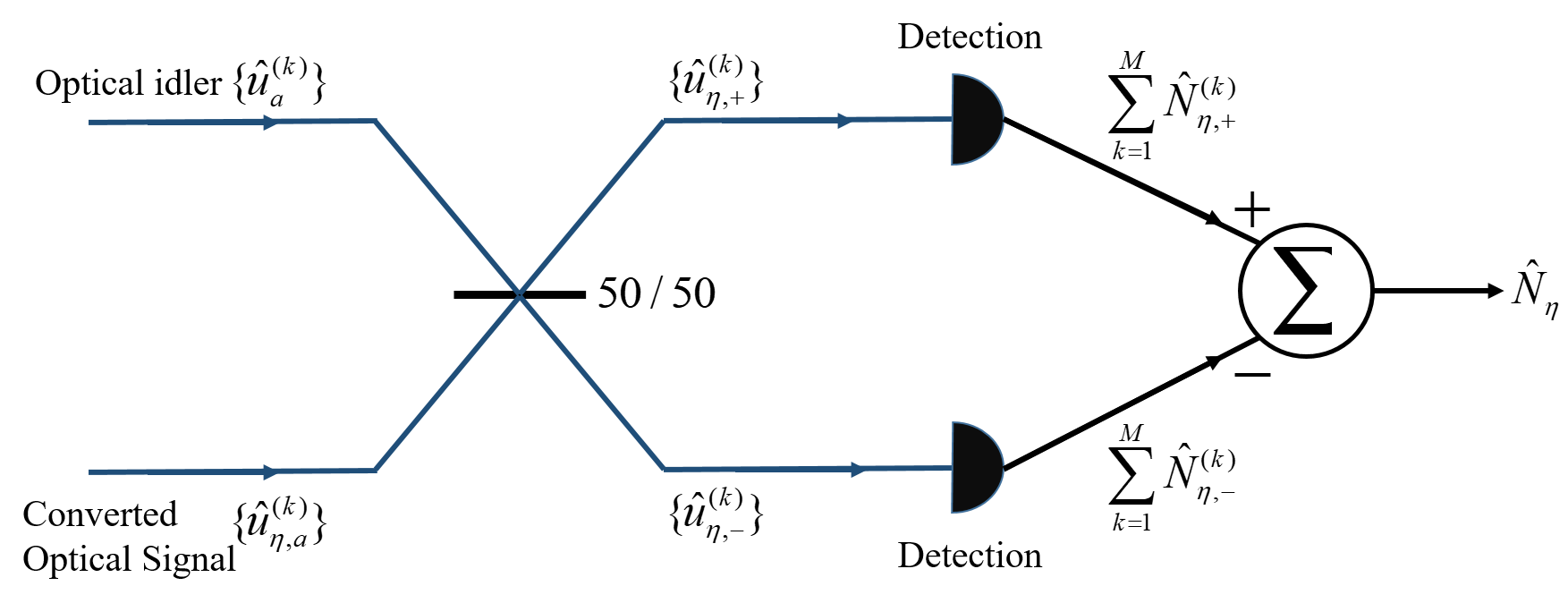}
	\caption{Optical post-processing of the QI receiver. The retained optical idler $\left \{ \hat{u}_{a}^{(k)} \right \}$ and the converted optical signal $\left \{ \hat{u}_{\eta, a}^{(k)} \right \}$ are mixed on a 50/50 beam splitter. The outputs of the beam splitter are detected, producing the measurements of $\sum\nolimits_{k=1}^{M} \hat{N}_{\eta, \pm}^{(k)}$, and the difference between $\hat{N}_{\eta, +}^{(k)}$ and $\hat{N}_{\eta, -}^{(k)}$ is used to quantum-enhanced decide whether the target is absent or present.}
	\label{fig:Fig1.png}
\end{figure}

The error probability has two parts, false-alarm probability $P_f \equiv \textup{Pr}$( $\textup{target present}$ | $\textup{target absent}$ $)$ and miss probability $P_m \equiv \textup{Pr}$( $\textup{target absent}$ | $\textup{target present}$ $)$. Since QI uses a large number of independent and identially distributed mode pairs, the Central Limit Theorem indicates that the measurement of $\hat{N}_{\eta}$ would produce a Gaussian random variable on target presence or absence. For equally-likely hypothesis, the error probability of the QI receiver satisfies \citep{S_ref8}
\begin{equation}
\begin{split}
P_{\textup{QI}}^{\textup{M}} = \frac{P_f + P_m}{2} = \frac{\textup{erfc} \left ( \sqrt{\textup{SNR}_{\textup{QI}}^{\textup{M}} / 8} \right )}{2},
\end{split}
\end{equation}
in which the $M$-mode signal-to-noise ratio is 
\begin{equation}
\begin{split}
\textup{SNR}_{\textup{QI}}^{\textup{M}} = \frac{4\left (\left \langle \hat{N}_{\eta} \right \rangle_{H_1} - \left \langle \hat{N}_{\eta} \right \rangle_{H_0} \right )^2}{\left ( \sqrt{\left \langle \Delta\hat{N}_{\eta}^2 \right \rangle_{H_0}} + \sqrt{\left \langle \Delta\hat{N}_{\eta}^2 \right \rangle_{H_1}} \right )^2},
\end{split}
\end{equation}
where $\left \langle \hat{N}_{\eta} \right \rangle_{H_j}$ and $\left \langle \Delta\hat{N}_{\eta}^2 \right \rangle_{H_j}$ are the conditional means and conditional variances of $\hat{N}_{\eta}$.

Due to the independent and identically distributed nature of the number operator $\hat{N}_{\eta, \pm }^{(k)}$, we can rewrite $\textup{SNR}_{\textup{QI}}^{\textup{M}}$ in terms of the single-mode moments, given by \citep{S_ref8}
\begin{equation}
\begin{split}
\textup{SNR}_{\textup{QI}}^{\textup{M}} = \frac{4 M \left [ \left ( \left \langle \hat{N}_{\eta,+} \right \rangle_{H_1} - \left \langle \hat{N}_{\eta,-} \right \rangle_{H_1} \right ) - \left ( \left \langle \hat{N}_{\eta,+} \right \rangle_{H_0} - \left \langle \hat{N}_{\eta,-} \right \rangle_{H_0} \right ) \right ]^2}{\left ( \sqrt{\left \langle (\Delta\hat{N}_{\eta,+}-\Delta\hat{N}_{\eta,-})^2 \right \rangle_{H_0}} + \sqrt{\left \langle (\Delta\hat{N}_{\eta,+}-\Delta\hat{N}_{\eta,-})^2 \right \rangle_{H_1}} \right )^2},
\end{split}
\end{equation}
where we have dropped the superscript $k$. With the help of Eq.(S5), Eq.(S10) and Eq.(S12), it is easily to obtain the terms that needed to calculate Eq.(S31)
\begin{equation}
\begin{split}
\left \langle \hat{N}_{\eta, \pm} \right \rangle_{H_0} = |B|^2[(\bar{n}_T + \bar{n}_b^T)/2 + 1] + |A_a|^2\bar{n}_a^T + |C_a|^2(\bar{n}_m^T + 1),
\end{split}
\end{equation}
\begin{equation}
\begin{split}
\left \langle \hat{N}_{\eta, \pm} \right \rangle_{H_1} = &\left \langle \hat{N}_{\eta, \pm} \right \rangle_{H_0} + \eta |B|^2[ |A_b|^2(\bar{n}_b^T+1) + |B|^2\bar{n}_a^T + |C_b|^2(\bar{n}_m^T+1) ]/2\\
&\pm \sqrt{\eta}\textup{Re}[|B|^2 A_b (\bar{n}_b^T + 1) - |B|^2 A_a \bar{n}_a^T + B^{*} C_a C_b (\bar{n}_m^T + 1)].
\end{split}
\end{equation}
For the variance, we have \citep{S_ref21}
\begin{equation}
\begin{split}
\left \langle (\Delta\hat{N}_{\eta,+}-\Delta\hat{N}_{\eta,-})^2 \right \rangle_{H_j} =& \left \langle \hat{N}_{\eta, +} \right \rangle_{H_j} \left (\left \langle \hat{N}_{\eta, +} \right \rangle_{H_j} + 1 \right ) + \left \langle \hat{N}_{\eta, -} \right \rangle_{H_j} \left (\left \langle \hat{N}_{\eta, -} \right \rangle_{H_j} + 1 \right )\\
&-\left ( \left \langle \hat{u}_{\eta,a}^{(k)\dagger} \hat{u}_{\eta,a}^{(k)} \right \rangle_{H_j} - \left \langle \hat{u}_{a}^{(k)\dagger} \hat{u}_{a}^{(k)} \right \rangle \right )^2/ 2,
\end{split}
\end{equation}
for $j = 0,1$, in which 
\begin{equation}
\begin{split}
\left \langle \hat{u}_{\eta,a}^{(k)\dagger} \hat{u}_{\eta,a}^{(k)} \right \rangle_{H_0} = \left | B \right |^2 (\bar{n}_T + 1) + \left | A_a \right |^2 \bar{n}_a^T + \left | C_a \right |^2 (\bar{n}_m^T + 1),
\end{split}
\end{equation}
\begin{equation}
\begin{split}
\left \langle \hat{u}_{\eta,a}^{(k)\dagger} \hat{u}_{\eta,a}^{(k)} \right \rangle_{H_1} = \left \langle \hat{u}_{\eta,a}^{(k)\dagger} \hat{u}_{\eta,a}^{(k)} \right \rangle_{H_0} + \eta |B|^2 [|A_b|^2(\bar{n}_b^T+1)+|B|^2 \bar{n}_a^T + |C_b|^2(\bar{n}_m^T+1)].
\end{split}
\end{equation}
Note that we assume that the target of interest is a low-reflectivity object, i.e., $\eta \ll 1$. Therefore, we have used the approximation $\bar{n}_T / (1-\eta) \approx \bar{n}_T$ in the former calculation.

\bibliography{Sup_Microwave.bib}